\documentclass[final,5p,times,twocolumn]{elsarticle}

\usepackage{amsmath,amssymb}
\usepackage{graphicx}

\begin{document}

\begin{frontmatter}

\title{Recent Progress  in the Theory  of the Crystalline Undulator}

\author{A.~Kostyuk}
\ead{andriy.p.kostyuk@gmail.com}

\address{Autogenstra{\ss}e 11, 65933 Frankfurt am Main, Germany}

\begin{abstract}
If an ultrarelativistic charged particle
channels inside a single crystal with periodically bent 
crystallographic planes, it 
emits hard electromagnetic radiation of the undulator
type. 
Due to similarity of its physical principles
to the ordinary (magnetic) undulator,
such a device is termed as the crystalline undulator.
Recent development of a new Monte Carlo code ChaS made possible a
detailed simulation of particle channeling and radiation emission 
in periodically bent crystals.
According to recent findings, energy of the electron beam 
below 1 GeV is sufficient to observe the undulator
effect in a periodically bent crystal. 
Even more exciting results were obtained for 
a crystalline undulator whose 
bending period is shorter than the period of the channeling 
oscillations and the
bending amplitude is smaller than the width of the planar channel.
Such a crystalline undulator is far superior to
what was proposed previously. 
It allows for a large effective number of 
undulator periods.
Therefore, it is
predicted to emit intense undulator radiation 
in the forward direction.
A narrow undulator peak is seen for both positron and electron beams.
Using positrons is, however,  more desirable because in this case the 
intensity of the undulator radiation is higher while 
the background is lower.
\end{abstract}

\begin{keyword}
channeling 
\sep 
radiation by moving charges 
\sep 
crystalline undulator 
\sep 
channeling radiation 
\sep 
electron and positron beams 
\sep 
synchrotron radiation sources 
\sep 
Monte Carlo method 
\sep
strained layer superlattice 

\PACS 
61.85.+p 
\sep 
41.60.-m 
\sep 
41.75.Fr 
\sep
02.70.Uu 



\end{keyword}

\end{frontmatter}

The idea of the crystalline undulator (CU) was proposed more than three decades
ago \cite{Kaplin1979,Kaplin1980,Baryshevsky1979,Baryshevsky1980}. It 
was suggested to use a periodically
bent crystal to generate hard electromagnetic radiation of undulator type.
Ultrarelativistic electrons or positrons were supposed to channel through such
a crystal following the sinusoidal shape of the bent crystallographic
planes. Due to nearly harmonic transverse oscillations of the particles,
the electromagnetic waves emitted in the forward direction
were expected to have a narrow spectral distribution, similarly to 
an ordinary (magnetic) undulator \cite{Ginzburg1947,Motz1951,Motz1953}.

The extremely strong electromagnetic fields inside the crystal can steer the 
beam particles much more effectively than it would be possible even with
the best superconductive magnets. Therefore, the period of the crystalline
undulator can be made orders of magnitude smaller than that of the magnetic
undulator. Hence, the crystalline undulator is potentially capable of
generating hard X rays and gamma rays.

There is, however, a price to pay for taking advantage of the crystalline field.
In contrast to the ordinary undulator, the particles move in a dense medium
instead of vacuum. They experience random collisions with crystal 
constituents and, therefore, they emit bremsstrahlung 
which may constitute a substantial background.
Coherent effects contaminate the spectrum even stronger. 
In addition to undulator oscillations, the particle has to 
perform channeling oscillations around the minimum of the planar potential 
(similarly to its motion in a straight crystal).
As a result, channeling radiation is present in 
the spectrum of the crystalline undulator in addition to undulator radiation and 
bremsstrahlung.

Channeling radiation has a lot in common with undulator radiation 
\cite{Kumakhov:1976ti}.
Therefore, even a straight crystal can be used to build a source of hard 
photons. However, this approach has a disadvantage. Because the shape of 
the transverse potential is not parabolic, the transverse motion of the 
particles is not harmonic. As a result, the spectrum of the channeling 
radiation is broader than that of the undulator radiation, 
especially in the case of negatively charged projectiles.\footnote{
Another difference between channeling radiation and undulator
radiation is that the latter one can be coherent if the particle
beam is modulated (bunched) in the longitudinal direction with the 
period equal to the wavelength of the undulator peak. In this case
the different particles radiate electromagnetic waves with nearly
the same phase. Therefore, the intensity of the radiation becomes 
proportional to the number of particles squared (in contrast to a 
linear proportionality for an unmodulated beam). This way of producing 
coherent radiation is utilized in free electron lasers \cite{Madey1971} 
(see e.g. \cite{Schmueser2008book} for a modern review). Similar 
effect should be observed in a crystalline undulator if it is fed 
by a bunched particle beam \cite{Kostyuk:2009js,Kostyuk_Patent}.
In contrast, channeling radiation will not become coherent even 
in the case of a modulated beam because the phases of the channeling 
oscillations are arbitrary and random.}

An interesting phenomenon of narrowing the planar channeling 
radiation peak has been observed for positrons 
\cite{Atkinson:1982bd,Bak:1984bk}.
The reason is a partial compensation of the potential inharmonicity
by deviations from the dipole approximation in the radiation
emission \cite{Ellison:1982zza}. This takes, however, place only
at a certain 'magic' beam energy. This means that the position of the 
narrow peak is fixed for every crystal channel, i.e. the radiation
frequency can be varied only within a very limited interval.

The CU approach does not have such restrictions. The undulator
radiation peak can be shifted within a wide range of values
by varying the beam energy and the bending period. 
But one has to put up with the contamination of the spectrum
by channeling radiation unless one finds a way
to get rid of it.

The radiation background is not the only challenge faced 
by the CU. It was realized at the very beginning 
\cite{Baryshevsky1980} 
that the effective length of the crystalline undulator 
is limited by the attenuation of photons in the crystal medium.
This limitation is, however, essential if rather soft 
photons, $\hbar \omega \lesssim 100$ keV are to be produced.

A more severe restriction on the effective length of the 
bent crystal is imposed by the dechanneling phenomenon 
\cite{Korol:1999im}.
The incoherent bremsstrahlung is not the only undesirable
effect of random scattering of the projectiles by the 
crystal constituents. Due to the random collisions,
the transverse energy of the channeling particle
fluctuates. Positive fluctuations are, however, more
likely than negative ones. Therefore, the projectile gains 
on average the transverse energy. If the 
latter exceeds the height of the interchannel potential barrier, 
the particle leaves the channel \cite{Lindhard1965}. 
Starting from this point, it does 
not follow the shape of the channel and, consequently, it does not emit undulator radiation. 
For this reason, the effective number of undulator periods $N_\mathrm{u}$ is 
limited by the average length $L_\mathrm{d}$ at which the dechanneling takes place:
\begin{equation}
N_\mathrm{u} \simeq \frac{L_\mathrm{d}}{\lambda_\mathrm{u}},
\label{Nu_Ld}
\end{equation}
where $\lambda_\mathrm{u}$ is the bending period of the crystalline undulator.
The number of undulator periods has to be large, $N_\mathrm{u} \gg 1$,
to ensure a narrow spectral distribution of the undulator radiation.
Therefore, the following condition has to be satisfied by the undulator 
period
\begin{equation}
\lambda_\mathrm{u} \ll L_\mathrm{d}.
\label{lambdau_Ld}
\end{equation}

In its initial form, the idea of the CU was based on the assumption that the projectile 
should follow the sinusoidal shape of the bent crystallographic planes and perform, at the 
same time, channeling oscillations around the central plane of the channel. This implied that 
the bending period of the undulator had to be much larger than the period of channeling oscillations
\begin{equation}
\lambda_\mathrm{u} \gg \lambda_\mathrm{c}.
\label{large_lambdau}
\end{equation}

In addition, the undulator bending amplitude $a_\mathrm{u}$ has to be 
much larger 
than the typical amplitude of channeling oscillations $a_\mathrm{c}$.
This condition can to be rewritten in the form \cite{Korol:1998ga}
\begin{equation}
a_\mathrm{u} \gg d
\label{large_a}
\end{equation}
due to the fact that $a_\mathrm{c} \lesssim d/2$, where $d$ is
the channel width (the distance between
the bent crystallographic planes that form the channel). 
The strong inequality (\ref{large_a}) has to be satisfied
to ensure a higher intensity of the undulator radiation 
relative to the channeling radiation.

In the following, the crystalline undulator satisfying conditions
(\ref{large_lambdau}) and (\ref{large_a}) will referred to as LALP CU 
(\textbf{l}arge \textbf{a}mplitude 
and \textbf{l}ong \textbf{p}eriod \textbf{c}rystalline \textbf{u}ndulator).
The complete list of conditions that have to be satisfied by the parameters of LALP CU
can be found in \cite{Korol:2004ug}.\footnote{It was argued in \cite{Korol:2004ug}
that condition (\ref{large_a}) is necessary to ensure separation of the undulator
radiation and the channeling radiation in the spectrum. In fact, it is not 
so. The separation is ensured by (\ref{large_lambdau}). The latter inequality
does follow from (\ref{large_a}) when it is supplemented by (\ref{stable_ch}), but 
(\ref{large_lambdau})
can be fulfilled even if $a_\mathrm{u} \lesssim d$. However, the intensity 
of the undulator radiation will be small in this case. This is the reason
why (\ref{large_a}) is necessary in the case of LALP CU.}

One more condition is relevant to the present discussion. It
ensures a stable channeling of the projectile in the periodically bent crystal
of LALP CU \cite{Kaplin1980,Baryshevsky1980}. Initially, similar condition
was obtained for channels with constant curvature \cite{Tsyganov1976}.
It is convenient to write it down in the 
form\footnote{
The centrifugal parameter can be also expressed in the form 
$C=R_\mathrm{c}/R_{\min}$, where $R_\mathrm{c}=E/U'_{\max}$ in the critical 
radius of the channel \cite{Tsyganov1976} (also known as Tsyganov's radius) and 
$R_{\min}=\lambda_\mathrm{u}^{2}/(4 \pi^{2} a_\mathrm{u})$ is the minimal curvature
radius of a sinusoid with the period $\lambda_\mathrm{u}$ and 
the amplitude $a_\mathrm{u}$.  
} 	
\begin{equation}
1 > C \equiv \frac{F_\mathrm{cf}}{U'_{\max}} =
4 \pi^{2} \frac{a_\mathrm{u} E}{\lambda_\mathrm{u}^{2} U'_{\max}} .
\label{stable_ch}
\end{equation}
Here $C$ is the centrifugal parameter \cite{Korol:2004ug},
$F_{\rm cf}$ is the maximal centrifugal force acting on the 
projectile in a the periodically bent channel, $U'_{\max}$
is the maximal force that keeps the particle in the channel and
$E$ is the energy of the projectile.

\begin{figure}[bt]
\includegraphics[width=0.95\linewidth]{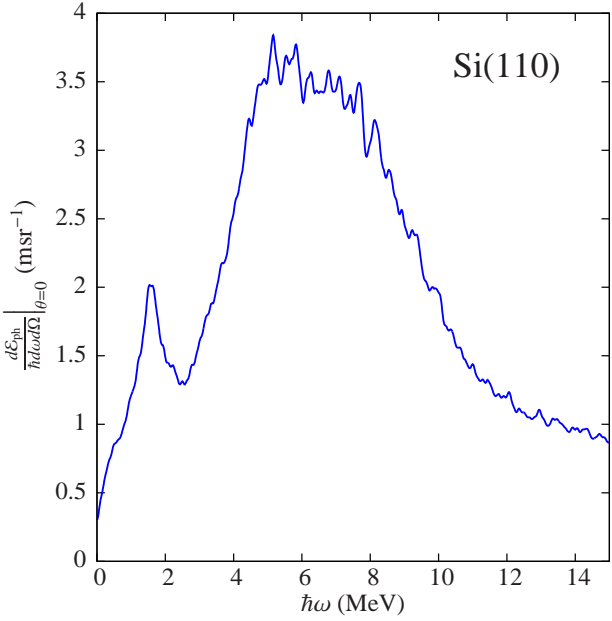}
\caption{Simulated spectrum of the radiation emitted by $E=855$ MeV
electrons channeling 
in a $24$ $\mu$m long LALP CU
with the bending period 
$\lambda_\mathrm{u}=4$ $\mu$m and the centrifugal
parameter $C=0.3$. The undulator radiation peak
is seen at $\hbar \omega \approx 1.6$ MeV. 
The spectrum is dominated by channeling radiation
(the broad maximum at $5$ MeV $\lesssim \hbar \omega \lesssim 9$ MeV ).
\label{e-_spectra_lalp}}
\end{figure}

Conditions (\ref{lambdau_Ld}), (\ref{large_a}) and (\ref{stable_ch})
are difficult to satisfy simultaneously. In particular, they cannot 
be satisfied in the case of
electron beam of moderate energy, $E \lesssim 1$ GeV.
For instance, if one sticks with (\ref{large_lambdau}),  
the most favorable conditions for observing
undulator radiation from $E = 855$ MeV electrons
in the Si(110) channel
are $\lambda_\mathrm{u} \approx 4$ $\mu$m and $C \approx 0.3$.
This corresponds to $a_\mathrm{u} \approx 0.84$ \AA\ 
($a_\mathrm{u}/d \approx 0.44$),
i.e. condition (\ref{large_a}) is broken. Taking into account
that the dechanneling length $L_\mathrm{d} \approx 8.3$ $\mu$m
even for the straight Si(110) channel 
\cite{Kostyuk:2010hs},\footnote{Dechanneling is defined in
\cite{Kostyuk:2010hs} as crossing the channel boundary.
Other authors (see e.g. \cite{BiryukovChesnokovKotovBook}) define
dechanneling as rising the energy of the channeling oscillations 
above the interchannel barrier. The value of the dechanneling length 
depends only slightly on the definition that is used for its computation.
The difference does not exceed a few hundreds of nanometers.}
one sees that the strong inequality (\ref{lambdau_Ld}) is not
satisfied either. As a result, the undulator peak is rather
small and not very sharp while the spectrum is dominated by the 
channeling radiation (see figure \ref{e-_spectra_lalp}).

Still, it is remarkable that the undulator effect is predicted
to be detectable
even using electron beams of moderate energies. This awakes expectation
that a successful proof of principle experiment can be done in the nearest
future, for example, at Mainz Microtrone (MAMI).

The LALP CU conditions can be fulfilled
for positron beams (see \cite{Korol:2004ug} and references therein)
and for high energy, $E > 10$ GeV, electrons
\cite{Tabrizi:2006yi}.

\begin{figure}[bt]
\includegraphics[width=0.95\linewidth]{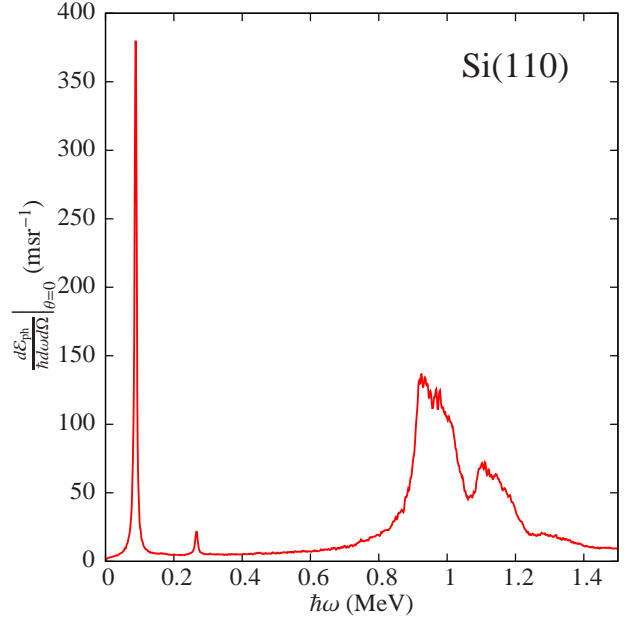}
\caption{Simulated spectrum of the radiation emitted by $E=500$ MeV
positrons channeling 
in a $350$ $\mu$m long LALP CU with the bending period 
$\lambda_\mathrm{u}=23.3$ $\mu$m and the centrifugal
parameter $C \approx 0.12$ 
($a_\mathrm{u}$=19.2 \AA{}, $a_\mathrm{u}/d \approx 10$). 
The undulator radiation peak
is seen at $\hbar \omega \approx 88$ keV. 
The parameters of the crystalline undulator are the
same as in right bottom panel of figure 8 of \cite{Korol:2004ug}.
In contrast to \cite{Korol:2004ug}, the present simulations
take into account the incoherent background.  
\label{e+_spectra_lalp}}
\end{figure}

An example of a spectrum of a positron-based LALP CU
is shown in figure \ref{e+_spectra_lalp}. The
undulator peak is narrow and is substantially
higher than the channeling radiation maximum.
Nevertheless, the total energy of  channeling radiation
(integrated over the photon energy interval
$0.8$ MeV $\lesssim \hbar \omega \lesssim 1.2$ MeV)
exceeds that of undulator radiation.
This is even more true for high energy electrons 
(see e.g.  figure 4 of \cite{Tabrizi:2006yi}). 
Moreover, the channeling photons are harder than the undulator ones
and, therefore, they cannot be easily screened out. 
This may cause serious problems for many potential applications.

\vspace{5mm}

The results of Monte Carlo simulations
shown in figures \ref{e-_spectra_lalp} and \ref{e+_spectra_lalp}
were obtained with a new  computer
code ChaS ({\bf Cha}nneling {\bf S}imulator). The code performs 
a 3D simulation of particle motion in the crystal and calculates the
spectral and angular distribution of the emitted radiation.

The previously existing channeling codes can be divided into two main categories.
There are codes based on the continuous potential approximation
\cite{Artru:1990nz,Biryukov:1995hv,Bogdanov_Mathematica_2010,Guidi_2010,Korol:2001ir,Saitoh_code_1985,Taratin_code_1979}
and those considering binary collisions of the projectile with the crystal atoms
\cite{Robinson1963,Kudrin_code_1973,Andersen:1979qs,Bak:1984fw,Smulders_code_1987}.
In the latter case, the atom is usually taken as as a whole, i.e. 
incoherent collisions with atomic electrons are ignored.
In contrast, the algorithm of ChaS is based on the binary collisions of the projectile with target 
electrons and target nuclei.
This novel 
approach is especially beneficial in the case of negatively charged projectiles, that 
have to cross the crystal plane during the channeling process. Indeed, the continuous potential approximation 
becomes inaccurate in the vicinity of the atomic nuclei while the electron density near the crystal 
plane is much higher than the average one.

The previous version of the code used the electron distribution calculated
from Moli\`ere's parametrization of the atomic potential. The obtained 
results were published in \cite{Kostyuk:2010hs,Kostyuk:2011kh}. They demonstrated 
reasonable agreement with experimental data \cite{Kostyuk:2010hs}. 
The present version of ChaS employs a more efficient and robust algorithm
for the calculation of the emitted radiation. 
In addition, it 
has an option of using the first principle
distribution of electrons in the crystal instead of Moli\`ere's parametrization. 
The first principle distribution  is calculated within the density functional 
theory using the computer code ABINIT \cite{abinit}.

The code ChaS in its present version ist most suitable 
for the analysis of the channeling of electrons
and positrons with energy $E$ in the range from a few hundreds of MeV to a few GeV
with the emition of not very soft: $\hbar \omega \gtrsim 0.5$ MeV
and at the same time not too hard $\hbar \omega \ll E$ photons.
The algorithm takes into account all the physics that is relevant 
to this domain. A number 
phenomena are neglected in the model (e.g. 
influence of the crystal medium on the emission and propagation of the radiation,
quantum effects in the motion of the projectile,
losses of the projectile energy due to emission of photons,
a shift of the photon energy due to recoil etc.)
They are expected to be small and do not influence the results substantially 
\cite{Kostyuk:2010hs,Uggerhoj:2005ms}.

For simplicity, the emittance of the particle beam 
was neglected in all simulations presented in this article.
The particles were assumed to enter the crystal at zero 
angle to the crystallographic planes.
 This is a reasonable approximation in the case 
of channeling experiments with high quality electron beams \cite{Backe:2008zz}.
It may not be the case for position beams, but the beam divergence depends
on details of the experimental conditions whose analysis is out of the scope
of the present contribution.

\vspace{5mm}

It was suggested recently \cite{Kostyuk:2012mk} that 
conditions (\ref{large_a}) and (\ref{stable_ch}) are not necessary.
In fact, an intense source of hard photons
with a narrow spectral distribution can can be created 
if both conditions are violated.

First, let us reanalyze the reasons behind condition (\ref{large_a}). 
It is needed to make sure that the spectrum is dominated
by the undulator radiation rather than by the channeling one. 
However, the amplitude of undulator oscillations has to be much larger
than that of the channeling oscillations
only if the frequency of the undulator radiation $\omega_\mathrm{u}$ is
smaller than the frequency of channeling oscillations $\omega_\mathrm{c}$.
Indeed, the energy radiated in a certain direction (the forward direction
in the present case) by a moving particle in the dipole approximation
has the following dependence on the transverse
oscillation amplitude $a$ and the radiation frequency $\omega$:
\begin{equation}
\left.
\frac{d \mathcal{E}}{d \omega \, d \Omega} 
\right |_{\theta=0}
\sim a^2 \omega^4.
\label{omega4}
\end{equation}
Here $d \Omega$ is the differential of the solid angle
and $\theta$ is the angle between the direction of the 
radiation emission and the average direction of the particle
motion.\footnote{The dependence (\ref{omega4}) becomes obvious
from the proportionality of the radiated energy to the particle
acceleration squared. For transverse harmonic oscillations, 
the transverse acceleration is proportional to the 
oscillation frequency squared and linearly proportional to
the oscillation amplitude. Hence the dependence (\ref{omega4})
is obtained.
}
Therefore, condition (\ref{large_a}) is not necessary, i.e.
the amplitude of the undulator bending can be smaller than the 
channel width,
\begin{equation}
a_\mathrm{u} < d,
\label{small_a}
\end{equation}
if the frequency of the undulator radiation is considerably
larger than that of the channeling radiation
\begin{equation}
\omega_\mathrm{u} \gg  \omega_\mathrm{c}.
\label{large_omegau}
\end{equation}
To fulfill this condition,  the
period of the crystal bending $\lambda_\mathrm{u}$ has to be much smaller
than the smallest period of channeling oscillations $\lambda_\mathrm{c}$:
\begin{equation}
\lambda_\mathrm{u} \ll \lambda_\mathrm{c}.
\label{small_lambdau}
\end{equation}
The last inequality violates condition (\ref{stable_ch}).
This can be seen from the following consideration.
The period of the channeling oscillations
can be estimated by
\begin{equation}
\lambda_\mathrm{c} \simeq 2 \pi \sqrt{\frac{E}{U''(0)}},
\label{lambdac}
\end{equation}
where $U''(0)$ is the second derivative of the transverse
potential energy with respect to $y$ in the point of its minimum $y=0$
(the axis $y$ is perpendicular to
the channel boundaries).  
Taking into account that\footnote{Expressions (\ref{lambdac}) and (\ref{Upmax})
would be exact equalities in the case of parabolic potential. For a real potential, 
the second derivative is not constant and the maximum value of the force is reached 
at $|y|<d$, hence (\ref{lambdac}) is an approximate equality and (\ref{Upmax})
is an inequality.}
\begin{equation}
U'_{\max} \lesssim U''(0)  d
\label{Upmax}
\end{equation}
in combination with (\ref{lambdac}) and (\ref{small_lambdau})
one obtains from (\ref{stable_ch})
\begin{equation}
1 > C \gg  \frac{a_\mathrm{u}}{d}.
\label{Cad}
\end{equation}
The bending amplitude of the crystalline undulator $a_\mathrm{u}$ 
cannot be much smaller than the channel width $d$ otherwise 
it becomes comparable to (or even smaller than) the amplitude of thermal vibrations
of the atoms in the crystal. Clearly, the undulator effect will not be present in this case.
If $a_\mathrm{u}$ is comparable to $d$, the two inequalities of  (\ref{Cad}) become 
incompatible.
One might expect that it destroyes the undulator effect, but, 
fortunately, it does not. Condition (\ref{stable_ch}) is, in fact,
not applicable in the case of SASP CU (\textbf{s}mall \textbf{a}mplitude 
(\ref{small_a}) and \textbf{s}hort 
\textbf{p}eriod (\ref{small_lambdau}) \textbf{c}rystalline \textbf{u}ndulator).

\begin{figure}[tb]
\includegraphics[width=0.95\linewidth]{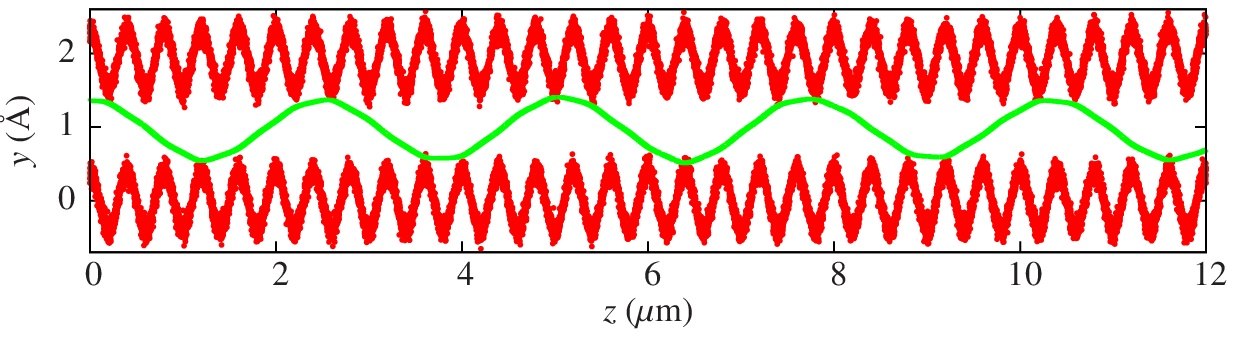}\\
\includegraphics[width=0.95\linewidth]{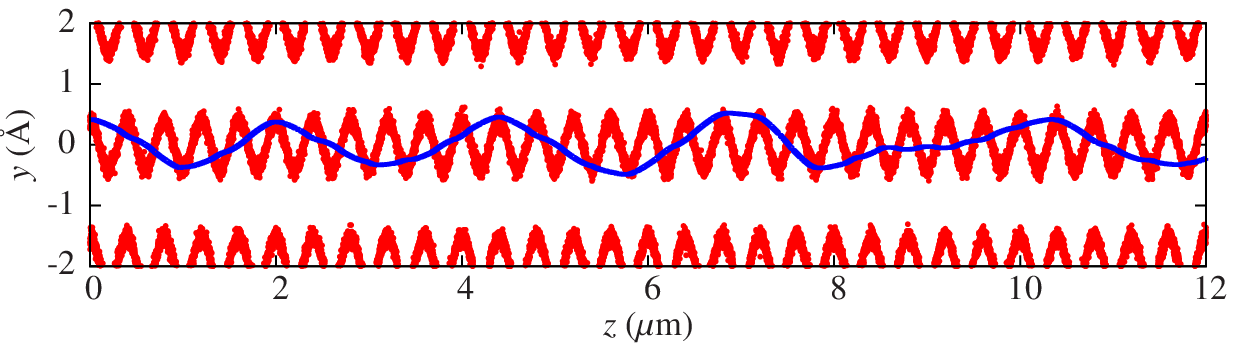}
\caption{Simulated trajectories of 
a positron (upper panel) and
an electron (lower panel) with energy $E=855$ MeV
channeling in a crystalline undulator
with a small amplitude, $a_\mathrm{u}=0.4$ \AA, and a short period, 
$\lambda_\mathrm{u}=400$ nm. The projectile does not 
follow the shape of the bent crystallographic planes (the thick wavy lines).
It performs channeling oscillations with roughly the same period 
as in a straight crystal. The effect of crystal bending on the shape of trajectories
is barely seen. The figure is a modified version of Fig. 1 published in 
\cite{Kostyuk:2012mk}.
\label{traj}}
\end{figure}

In figure \ref{traj}, simulated trajectories of a positron and  an 
electron channeling in a SASP CU  are plotted. 
One sees from the figure that the particles do not follow the
shape of the bent crystallographic planes. 
Therefore, formula (\ref{stable_ch}) is irrelevant.
The channeling process is still present.
The particle motion can be roughly considered as if it were
governed by a continuous potential averaged
over the oscillations of the plane. In other words, it is similar
to the channeling in a straight crystal but the average transverse 
potential is somewhat different.

\begin{figure}[bt]
\includegraphics[width=0.95\linewidth]{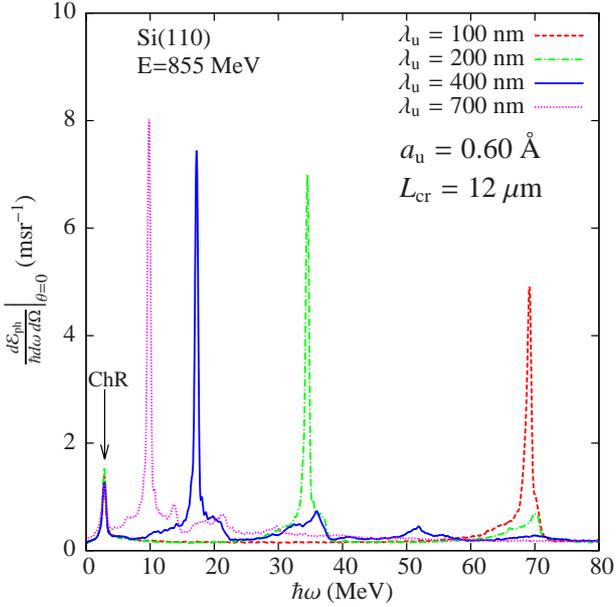}
\caption{Simulated spectra of the radiation emitted by $E=855$ MeV
positrons channeling 
in a $12$ $\mu$m long
crystalline undulator with a small bending amplitude for four
short periods. The spectra are obtained by averaging over about
1000 simulated trajectories.
The undulator radiation peaks
are higher
and are centered at a much larger photon energy
than the corresponding channeling radiation (ChR) peaks. 
\label{e+_spectra}}
\end{figure}

\begin{figure}[tb]
\includegraphics[width=0.95\linewidth]{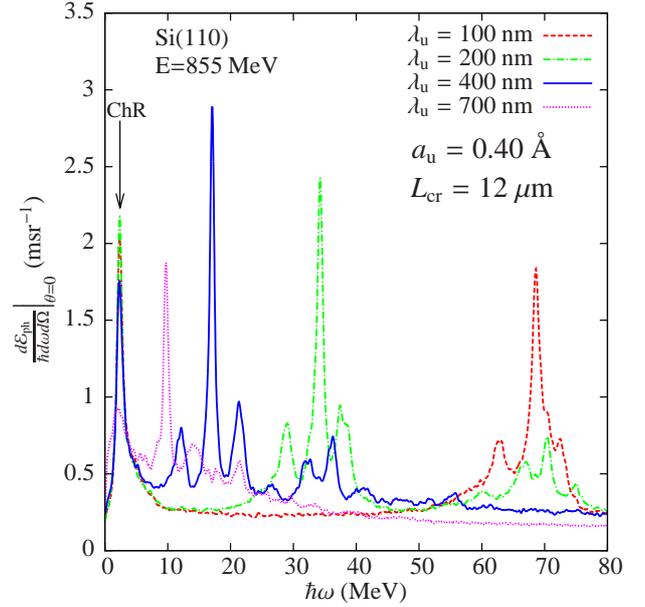}
\caption{The same as in figure \ref{e+_spectra}
but for electrons. The undulator radiation peaks
are lower than in figure \ref{e+_spectra}  by a factor 
of $\sim 3$ and the background is higher.
\label{e-_spectra}}
\end{figure}

Still, a more thorough consideration reveals
that the particle performs also transverse oscillations 
with the undulator period
$\lambda_\mathrm{u}$. The amplitude of 
these 
oscillations $a$ is much smaller than the bending amplitude 
$a_\mathrm{u}$.
Therefore, it is difficult  to see the
modification of the trajectories induced by the crystal 
bending in figure \ref{traj}.
Nonetheless, the corresponding Fourier harmonics are present 
and they are seen 
in the spectra of the emitted radiation shown in figures \ref{e+_spectra}
and \ref{e-_spectra}.

The undulator radiation peak 
is  higher than the channeling radiation maximum
despite of the amplitude $a$ of the undulator 
oscillations of the projectile being
much smaller than the amplitude of its channeling oscillations
$a_\mathrm{c}$.
This seemingly paradoxical fact can be easily
understood  if relations
(\ref{omega4}) and (\ref{large_omegau})
are taken into consideration.

The undulator peaks are narrow and well separated from the 
channeling radiation. The undulator and channeling 
radiation peaks have approximately the same
\textit{absolute} width. Nonetheless, 
the \textit{relative} width of undulator peaks
is much smaller due to (\ref{large_omegau}).

Due to small $\lambda_\mathrm{u}$, condition 
(\ref{Nu_Ld}) can be satisfied for SASP CU even 
if it is used with a moderate energy electron beam. This is 
a big practical advantage since electron 
beams are usually of higher quality and are less 
expensive than positron ones. Still, positrons
are more preferable. As one sees from 
figures \ref{e+_spectra} and \ref{e-_spectra},
they allow for a higher intensity 
of the undulator peak accompanied by a 
lower background.

The undulator radiation of SASP CU
is much harder than the channeling 
radiation (\ref{large_omegau}). It is an important 
advantage of SASP CU over LALP CU.
It is usually much easier to get rid of
a soft photon background and preserve the 
hard part of the spectrum than to do the opposite.
For example, a detector can be made 
sensitive to hard photons, but be screened from (or 
made insensitive to) soft photons. It is 
difficult and sometimes even  impossible 
to do vice versa. For this reason,  
SASP CU is expected to be much more suitable for 
many potential applications than LALP CU.

Due to its much smaller bending period, SASP CU
can produce by about two orders of magnitude harder 
photons when used with a beam of the same energy as
LALP CU. Or, in other words,  SASP CU will
require a much smaller and, therefore, a much
less expensive accelerator than the one which would 
be needed for the production 
of radiation of the same frequency with LALP CU.

Hence, the crystalline undulator that violates conditions 
(\ref{large_a}) and (\ref{stable_ch}) has a number of advantages 
with respect to LALP CU.

From the technological point of view, SASP CU is more challenging 
than LALP CU. There exist at least four technologies suitable
for the fabrication of LALP CU. The oldest idea of 
using ultrasonic waves \cite{Kaplin1980,Baryshevsky1980}
is, unfortunately, still waiting for its experimental implementation.
Two other technologies utilize the idea of imposing periodic 
stresses
on the surface of the crystal sample.
These are
making regularly spaced grooves on the crystal surface
either by a diamond blade \cite{Bellucci:2002dy,Guidi:2005ya}
or by means of laser-ablation \cite{Balling:2009zz} and
deposition
of periodic
Si$_3$N$_4$ layers onto the surface of a Si crystal \cite{Guidi:2005ya}.
Finally, there is a way to create periodically varying stresses in the
crystal volume by growing a crystal with periodically varying chemical composition.
The most mature technology is the creation of strained layer \cite{Breese:1997va}
superlattices by 
periodically varying germanium concentration $x$ in 
a Si$_{1-x}$Ge$_x$ crystal
with a periodically varying Ge content $x$ 
\cite{Mikkelsen:2000ky,Krause:2001ew}.

The potential applicability of ultrasound in the case of SASP CU
requires further investigations.
The methods based on surface stresses cannot be applied because
they require the transverse dimension of the crystal to be of the order
of the bending period \cite{Kostyuk:2007ny}. The latter is smaller 
than 1 $\mu$m in the case of SASP CU.
Only the last approach, the growing of Si$_{1-x}$Ge$_x$
strained layer superlattices,
is suitable for the fabrication of SASP CU. This technology has been 
already used for manufacturing the LALP CU that 
is being used in ongoing experiments at Mainz Microtron \cite{Backe2011}.
It was demonstrated recently \cite{Kostyuk:2013kk} that a Si$_{1-x}$Ge$_x$
crystalline heterostructure can be grown 
with the parameters that have been used in the simulations presented in 
figures \ref{e+_spectra} and \ref{e-_spectra}.
Moreover, the strained layer crystal with parameters of  SASP CU was predicted 
to be stable agains misfit dislocations, in contrast to LALP CU which is 
only metastable.

In conclusion, the crystalline undulator with a small amplitude and 
a short period can be created and it is predicted to be far 
superior with respect to LALP CU. 

In the present contribution, production of undulator
radiation with photon energy in the range of tens of 
megaelectronvolts is considered. These results are important
because the hard photon range is unattainable
for the present state-of-the-art synchrotron radiation sources.
It would be, however, interesting to consider production of 
softer photons, in the range of a few hundreds or even tens 
of kiloelectronvolts. This domain is on the edge of the 
capability of the presently existing and constructed facilities 
\cite{Schmueser2008book,Willmott2011book}.
This facilities are, however, unique and very expensive.
Due to the fact that SASP CU requires a much smaller accelerator
than a conventional undulator
for production of photons of the same energy, it has a potential
to offer a much less expensive solution.
If it were possible to use SASP CU for production of soft
X rays, such devices could be made affordable even to medium
size university labs or hospitals. 
A theoretical investigation of the low energy SASP CU is,
however, more challenging, because quantum effects definitely
cannot be neglected in this case.

Even more exciting future task of this field of science is 
exploring the possibility to produce
coherent radiation with SASP CU \cite{Kostyuk_Patent}. 

\section*{Acknowledgements}

I am very much obliged to the Organizing Committee of the 
International Conference ``Channeling 2012'' and especially 
to Sultan Dabagov for the great opportunity to present my
scientific results in Alghero.

The financial support of the European Physical Society
is highly appreciated.

I am grateful to
X. Artru, H. Backe, V. Baryshevsky, S. Dabagov, 
M. Garattini, V. Guidi, A. Mazzolari, M. Motapothula,
A. Shchagin, N. Shul'ga, 
V. Tikhomirov, W. Wagner for their interest to the work, 
for stimulating discussions and critical comments.

At its initial stage, the work was partially done in Frankfurt Institute
for Advanced Studies (FIAS) and was supported by
Deutsche Forschungsgemeinschaft (DFG).
Numerical simulations were done at the 
Center for Scientific Computing of 
the J.W.~Goethe University, 
Frankfurt am Main (Germany).


\begin{thebibliography}{99}
%
\bibitem{Kaplin1979}
V.~V.~Kaplin, S.~V.~Plotnikov, and S.~A.~Vorobiev, in Abstracts of the 10th Conference 
on the Application of Charged Particle Beams for Studying the Composition and Properties 
of Materials (Moscow State University, Moscow, 1979), p. 28.
%
\bibitem{Kaplin1980}
V.~V.~Kaplin, S.~V.~Plotnikov and S.~A.~Vorobiev, 
Zh.\ Tekh.\ Fiz.\ 50, 1079--1081 (1980), (Sov.
Phys. – Tech. Phys. 25, 650--651 (1980)).
%
\bibitem{Baryshevsky1979}
V.~G.~Baryshevsky, in 
Proceedings of the 14th Winter School on Physics of Atomic Nucleus
(Leningrad Inst. of Nucl. Phys., Leningrad,  1979), p. 158--167 (in Russian).
%
\bibitem{Baryshevsky1980}
V.~G.~Baryshevsky,  I.~Ya.~Dubovskaya  and  A.~O.~Grubich,  
Phys.\ Lett.\ A {\bf 77}, 61--64 (1980).
%
\bibitem{Ginzburg1947}
V.L.~Ginzburg, Izv. Akad. Nauk. SSSR, Ser. Fiz. {\bf 11}, 165--181 (1947).
%
\bibitem{Motz1951}
H. Motz, J. Appl. Phys. {\bf 22}, 527--535, ibid. 1217
(1951).
%
\bibitem{Motz1953} H. Motz, W. Thon, and R. N. Whitehurst
J. Appl. Phys. {\bf 24}, 826--833 (1953).
%
\bibitem{Kumakhov:1976ti} 
M.~A.~Kumakhov,
Phys.\ Lett.\ A {\bf 57}, 17--18 (1976).
%
\bibitem{Madey1971}
J.M.J.~Madey, J. Appl. Phys. {\bf 42}, 1906--1913 (1971).
%
\bibitem{Schmueser2008book}
P.~Schm{\"u}ser, M.~Dohlus, J.~Rossbach,
{\it ``Ultraviolet and Soft X-Ray Free-Electron Lasers''}, Springer, Berlin Heidelberg,
(2008).
%
\bibitem{Kostyuk:2009js}
A.~Kostyuk, A.~V.~Korol, A.~V.~Solov'yov and W.~Greiner,
J.\ Phys.\ B {\bf 43}, 151001 (2010).
%
\bibitem{Kostyuk_Patent}
W.~Greiner, A.~V.~Korol, A.~Kostyuk, A.~V.~Solov’yov, 
``Vorrichtung und Verfahren zur Erzeugung elektromagnetischer Strahlung'', 
German Patent DE102010023632, December 20 (2011).
%
\bibitem{Atkinson:1982bd}
M.~Atkinson {\it et al.},
Phys.\ Lett.\ B {\bf 110}, 162--166 (1982).
%
\bibitem{Bak:1984bk}
J.~Bak {\it et al.},
Nucl.\ Phys.\ B {\bf 254}, 491--527 (1985).
%
\bibitem{Ellison:1982zza}
J.~A.~Ellison {\it et al.},
Phys.\ Lett.\ B {\bf 112}, 83--87 (1982).
%
\bibitem{Korol:1999im} 
A.~V.~Korol, A.~V.~Solovov and W.~Greiner,
Int.\ J.\ Mod.\ Phys.\ E {\bf 8}, 49--100 (1999).
%
\bibitem{Lindhard1965}
J.~Lindhard,
Kong. Danske Vid. Selsk. Mat.-Fys. Medd. 34(14) 
 (1965).
%
\bibitem{Korol:1998ga} 
A.~V.~Korol, A.~V.~Solovov and W.~Greiner,
J.\ Phys.\ G {\bf 24}, L45--L53 (1998).
%
\bibitem{Korol:2004ug} 
A.~V.~Korol, A.~V.~Solov'yov and W.~Greiner,
Int.\ J.\ Mod.\ Phys.\ E {\bf 13}, 867--916 (2004).
%
\bibitem{Tsyganov1976}
E.N.~Tsyganov, TM-682, TM-684,
Fermilab, Batavia
(1976).
%
\bibitem{Kostyuk:2010hs} 
A.~Kostyuk et al.,
J.\ Phys.\ B.\ At.\ Mol.\ Opt.\ Phys.\  {\bf 44}, 075208 (2011).
%
\bibitem{BiryukovChesnokovKotovBook}
Biryukov~V~M,  Chesnokov~Yu~A, Kotov~V~I
1997
{\it Crystal Channelling and its Application at High-Energy  Accelerators}
(Berlin Heidelberg New York: Springer) 
%
\bibitem{Tabrizi:2006yi} 
M.~Tabrizi et al.,
Phys.\ Rev.\ Lett.\  {\bf 98}, 164801 (2007).
%
\bibitem{Artru:1990nz} 
X.~Artru,
Nucl.\ Instrum.\ Meth.\ B {\bf 48}, 278--282 (1990).
%
\bibitem{Biryukov:1995hv} 
V.~Biryukov,
Phys.\ Rev.\ E {\bf 51}, 3522--3528 (1995).
%
\bibitem{Bogdanov_Mathematica_2010}
V.O.~Bogdanov {\it et al.},
J.\ Phys.:\ Conf.\ Ser.\ {\bf 236}, 012029  (2010).
%
\bibitem{Guidi_2010}
V.~Guidi, A.~Mazzolari and V.~Tikhomirov,
J.\ Appl.\ Phys. {\bf 107}, 114908 (2010).
%
\bibitem{Korol:2001ir} 
A.~V.~Korol, A.~V.~Solovov and W.~Greiner,
J.\ Phys.\ G {\bf 27}, 95--125 (2001).
%
\bibitem{Saitoh_code_1985}
K. Saitoh, 
J.\ Phys.\ Soc.\ Japan, {\bf 54} 152--161 (1985).
%
\bibitem{Taratin_code_1979}
A.M.~Taratin, E.N.~Tsyganov and S.A.~Vorobiev, 
Phys.\ Lett.\ A {\bf 72} 145--146 (1979).
%
\bibitem{Robinson1963} 
M.~T.~Robinson and O.~S.~Oen,
Appl. Phys. Lett. {\bf 2}, 30--32  (1963);
Phys.\ Rev.\  {\bf 132}, 2385--2398 (1963).
%
\bibitem{Kudrin_code_1973}
V.V.~Kudrin, Yu.~A. Timoshenkov and S.A.~Vorobiev, 
Phys.\ Stat.\ Sol.\ B {\bf 58} 409--414 (1973).
%
\bibitem{Andersen:1979qs} 
S.~K.~Andersen {\it et al.},
Nucl.\ Phys.\ B {\bf 167}, 1--40 (1980).
%
\bibitem{Bak:1984fw} 
J.~F.~Bak {\it et al.},
Nucl.\ Phys.\ B {\bf 242}, 1--30 (1984).
%
\bibitem{Smulders_code_1987}
P.J.M.~Smulders and D.O.~Boerma,
Nucl. Instrum.  Meth. B {\bf 29}, 471--481 (1987).
%
\bibitem{Kostyuk:2011kh}
A.~Kostyuk et al.,
arXiv:1104.3890 [physics.acc-ph].
%
\bibitem{abinit}
X.~Gonze {\it et al.},
Zeit.\ Kristallogr.\ {\bf 220}, 558--562 (2005);
Comp.\ Phys.\ Commun. {\bf 180}, 2582--2615 (2009).
%
\bibitem{Uggerhoj:2005ms} 
U.~I.~Uggerh{\o}j,
Rev.\ Mod.\ Phys.\  {\bf 77}, 1131--1171 (2005).
%
\bibitem{Backe:2008zz}
H.~Backe {\it et al.}
Nucl.\ Instrum.\ Meth.\ B {\bf 266}, 3835--3851 (2008).
%
\bibitem{Kostyuk:2012mk}
A.~Kostyuk,
Phys. Rev. Lett. (in print) [arXiv:1210.0468]. 
%
\bibitem{Bellucci:2002dy}
S.~Bellucci {\it et al.},
Phys.\ Rev.\ Lett.\  {\bf 90}, 034801 (2003).
%
\bibitem{Guidi:2005ya}
V.~Guidi {\it et al.},
Nucl.\ Instrum.\ Meth.\ B {\bf 234}, 40--46 (2005).
%
\bibitem{Balling:2009zz}
P.~Balling {\it et al.},
Nucl.\ Instrum.\ Meth.\ B {\bf 267}, 2952--2957 (2009).
%
\bibitem{Breese:1997va} 
M.~B.~H.~Breese,
Nucl.\ Instrum.\ Meth.\ B {\bf 132}, 540--547 (1997).
%
\bibitem{Mikkelsen:2000ky} 
U.~I.~Uggerh{\o}j and E.~Uggerh{\o}j,
Nucl.\ Instrum.\ Meth.\ B {\bf 160}, 435--439 (2000).
%
\bibitem{Krause:2001ew}
W.~Krause {\it et al.},
Nucl.\ Instrum.\ Meth.\ A {\bf 483}, 455--460 (2002).
%
\bibitem{Kostyuk:2007ny}
A.~Kostyuk {\it et al.},
Nucl.\ Instrum.\ Meth.\ B {\bf 266}, 972--987 (2008).
%
\bibitem{Backe2011}
H.~Backe,
Nuovo Cimento C, {\bf 34}(4), 157--165 (2011).
%
\bibitem{Kostyuk:2013kk}
A.~Kostyuk,
arXiv:1301.4491 [physics.acc-ph].
%
\bibitem{Willmott2011book} 
Philip~Willmott,
{\it ``An Introduction to Synchrotron Radiation: Techniques and Applications''},
John Wiley and Sons (2011). 
%
\end{thebibliography}
\end{document}